\documentclass[smallextended]{svjour3}
\smartqed
\usepackage{graphicx}
\usepackage{amsmath}
\usepackage{amssymb}
\newtheorem{Theorem}{Theorem}

\begin{document}
	
	\title{Measurement-Device-Independenization of Quantum Key Distribution Protocols}
	\author{Hao Shu}
	\institute{Hao Shu\at
		Shenzhen University
		\\
		South China University of Technology
	    \\
		\email{Hao\_B\_Shu@163.com}
	}
	
	\date{}	
	
	\maketitle
	
	\begin{abstract}
		Quantum key distribution(QKD) allows the legitimate partner to establish a secret key whose security only depends on physical laws. In recent years, research on QKD by employing insecure measurement devices, namely measurement-device-independent QKD (MDI-QKD) is increased. MDI-QKD removes all attacks on measurement devices and thus an untrusted third party can be employed for measuring. However, a weakness of previous MDI-QKD protocols is the need for joint measurements such as Bell measurements whose efficiency is low in practice. On the other hand, can all QKD protocols become measurement-device-independent remains a problem. In this paper, we present a scheme making prepare-measure QKD protocols become MDI-QKD protocols, called $'measurement-device-independenization'$, which does not need to employ joint measurements and could be efficiently implemented by weak coherence sources. The protocol might look like the detector-device-independent(DDI) protocols but it is also secure under the Trojan horse attack. To illustrate this, we investigate the photon-number-adding(PNA) attack and present a scheme, called $'photon-number-purification'$, which can also be employed to close loopholes for previous protocols such as DDI and plug-and-play ones.
		
	\end{abstract}
	
	\section{Introduction}
	
	Quantum key distribution(QKD) could be the most significant application in quantum information theory, which allows the legitimate partner to share a one-time pad with security that only depends on physical laws. In nearly four decades, substantial QKD protocols are proposed\cite{BB1984Quantum,E1991Quantum,BB1992Quantum,B1992Quantum,GV1995Quantum,B1998Optimal,LC1999Unconditional,CB2002Security,SA2004Quantum,K2006A,ST2016A,GR2010Quantum,S2021Quantum}. Traditional security analyses of QKD protocols focus on attacks on channels. However, the devices of the legitimate partner may also be insecure. To solve this problem, research on device-independent quantum key distribution(DI-QKD)\cite{E1991Quantum,BH2005No,AB2007Device,HR2010Device,MS2011Secure,LP2013Device,AR2016Completely} becomes interesting. Despiteness, completed DI-QKD usually results in impractical requirements. To balance the security and the practice, measurement-device-independent quantum key distribution(MDI-QKD) appears.
	
The setting of MDI-QKD is that QKD protocols are implemented by insecure measurement devices, namely it might send the measurement outcomes to an eavesdropper, might employ a different measurement instead of the required one, might provide a fake outcome, and so on. In the worst case, the measurement might be totally controlled by the eavesdropper who can do whatever can be done under physical laws to obtain the best interests. Therefore, an MDI-QKD can be simply understood as a QKD in which measurements are provided by an untrusted third party. The first MDI-QKD protocol\cite{LC2012Measurement} was published in 2012, in which respectively, the legitimate partner randomly chooses one of the two BB84 bases and sends one of the states randomly to an untrusted third party for measuring via a Bell measurement, for each bit. If their bases agree, they can know whether the states they prepared are the same or different. Since they know the bit they prepared, respectively, they can learn the other one and thus agree with a key, while the third party only knows whether they have the same bits but does not learn what they are, and thus their key is private for the party. After the protocol was published, several relative works followed\cite{MR2012Alternative,MT2014Measurement,JY2017Measurement,WT2021Measurement,DZ2021Measurement,JY2021Higher}. In 2018, the idea of twin-field QKD\cite{LY2018Overcoming} was proposed, which might overcome traditional limitations of distance\cite{PG2009Direct,PL2017Fundamental}. Inspirited by it, other QKD protocols appeared\cite{MZ2018Phase,CY2019Twin}.
	
	A weakness of all such MDI-QKD protocols is the need for joint measurements such as Bell measurements which are lowly efficient in practice, while a completed Bell measurement can not be provided by linear optics\cite{LC1999Bell}. Recently, an MDI-QKD scheme without joint measurements was presented\cite{H2021Measurement} by employing order-rearrangement technology. However, it still needs to be supplied by other technologies, for example, teleportations (which also result in the need for Bell measurements) as the author suggested, to obtain security. Therefore, as far as we know, no previous MDI-QKD protocols can avoid joint measurements securely. On the other hand, can all QKD protocols become measurement-device-independent ones remains a problem.
	
	In this paper, we present a scheme without employing joint measurements, making a prepare-measure QKD protocol measurement-device-independent, which would be called $'measurement-device-independenization'$ of QKD protocols, see Figure 1. The protocol might look like the detector-devices-independent(DDI) QKD protocols which are insecure under Trojan horse attacks\cite{LK2014Detector,Q2015Trustworthiness,GR2015Quantum,LL2015Simple,SH2016Insecurity}. However, after applying a new scheme, our protocol can be immune to the Trojan horse attack in DDI protocols. To illustrate this, we investigate the photon-number-adding(PNA) attack which is exactly the Trojan horse attack when applying to DDI protocols, and provide a solution called the photon-number-purification(PNP) scheme. The scheme can also close loopholes in previous protocols, such as the DDI protocol, the protocol in \cite{H2021Measurement}, and the plug-and-play protocol\cite{MH1997Plug,GF2006Trojan}. Finally, we provide a protocol under practical assumptions.
	
The main advances of the paper include (1) We demonstrate that Joint measurements can be avoided securely in an MDI-QKD protocol; (2) We present a scheme by which any prepare-measure QKD protocol can become an MDI-QKD protocol with nearly the same security and efficiency; (3) We present the PNP scheme to solve a kind of two-stage attack, which is considered as a Trojan horse attack in some previous protocols, in QKD.

\begin{figure}
	\includegraphics[width=1.0\textwidth]{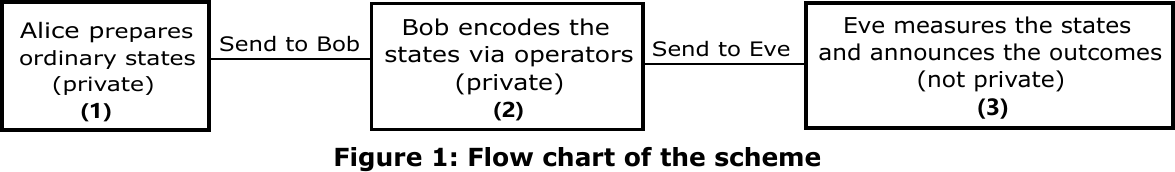}
\end{figure}

	\section{For BB84 type protocol under ideal assumptions}
	
	Let us investigate prepare-measure QKD protocols of BB84 type with assumptions that the legitimate partner, Alice and Bob, employs perfect devices except ones for measuring, for simplicity.
	
	We first state how employing insecure measurement devices can influence the BB84 protocol. In the ordinary BB84 protocol, the sender, Alice, chooses one of the two bases $\{|0\rangle, |1\rangle\}$ and $\{|+\rangle, |-\rangle\}$ randomly and sends one of the states in the chosen basis randomly to the receiver, Bob. After Bob receives the state, he randomly chooses one of the bases for measuring. If the choices of the bases of the legitimate partner are the same, they go to the next step, else discard the bit. There are a lot of attacks an eavesdropper, Eve, can employ if she controls the measurement devices. For example, she can intercept the state sent by Alice, measuring via a randomly chosen basis above, and resend the output state to Bob. If Bob chooses the different basis as she chose, then she lets the measurement devices report nothing (since she controls the devices), pretending that the state was lost. Hence, all effective states are eavesdropped on by Eve without improving any error as she measured the states sent by Alice with the same bases chosen by Alice. In fact, a simpler scenario can be that Eve reads the measurement outcomes and obtains the same information as Bob if she controls the detectors.
	
	The explanation above demonstrates that the BB84 protocol is not an MDI-QKD protocol. However, employing a novel scheme, a BB84-type protocol that can be implemented by insecure measurement devices is provided as follow.
\newpage
	
	\textbf{BB84 type Protocol:}

	\textbf{Step 1 (Alice's turn)}: Alice prepares a state for each bit randomly in one of the four states $|0\rangle, |1\rangle, |+\rangle, |-\rangle$ and sends it to Bob. Alice and Bob agree that the legitimate bit is 0 if Bob encodes the state by Pauli operator $Z$ or Hadamard gate $H$ and 1 if Bob encodes by Pauli operator $X$ or operator $HXZ$.
	
	\textbf{Step 2 (Bob's turn)}: After receiving the state, Bob implements the PNP scheme firstly (see section VI), and then provides a unitary operator, randomly in $Z$, $X$, $H$ and $HXZ$ on the state employed for coding. Then he sends the employed state to the untrusted third party, Eve, for measuring.
	
	\textbf{Step 3 (Eve's turn)}: Eve is required to measure the state via basis $\{|0\rangle, |1\rangle\}$, or $\{|+\rangle, |-\rangle\}$, randomly (the basis can be chosen by Alice or Bob to improve the efficiency) and publicly announce the outcome.
	
	These steps will be repeated several times until Alice and Bob share a long enough bit string.
	
	\textbf{Step 4 (sifting turn)}: For each bit, Alice publicly announces the chosen basis as in BB84 protocol while Bob announces whether his operator is in $\{H, HXZ\}$ (but not what it is). If Bob's operator belongs to $\{H, HXZ\}$ while the basis of Alice and Eve are the same or if Bob's operator does not belong to $\{H, HXZ\}$ while the bases of Alice and Eve are different, the bit will be discarded.
	
	\textbf{Step 5 (error estimating and raw key generating turn)}: Bob chooses part of the bit string for error estimating. He announces his operations on those bits for Alice who knows the ordinary states to estimate the bit error rate. If the error rate is acceptable, they generate a raw key by the remaining bits and continue error-correcting and privacy amplification procedures.
	\\
	
	The protocol is MDI because the whole measurement is implemented by an untrusted third party Eve while the legitimate partner does not need to know if Eve is honest. Note that in the protocol, Bob knows what he encoded by while Alice knows the states she sent as well as the measurement outcomes, and therefore, also knows what Bob encoded by. On the other hand, however, since Eve only knows the measurement outcomes, the legitimate bits are random for her. Also, note that the protocol might look like the DDI protocols which are insecure under Trojan horse attacks. However, they are different. On one hand, the above protocol is measurement-device independent, not only detector-device independent, since the measurement is assumed to be implemented by an untrusted third party. On the other hand, the above one can be immune to the Trojan horse attack, thanks to the PNP scheme (which will be discussed in section VI). From now on, let us ignore the Trojan horse attack temporarily until section VI.
	
	\section{Security and efficiency}
	
	The security of the protocol comes from the security of the ordinary BB84 protocol and the indistinguishability of operators employed by Bob. Precisely, to access a bit, the eavesdropper, without loss generality says Eve, which is the worst case for Alice and Bob, has to discriminate either the state sent by Alice or the operator provided by Bob. The security of the first case is the same as in the ordinary BB84 protocol while the security of the second one comes from the indistinguishability of the operators employed by Bob.
	
	In detail, $X$, $Z$, $H$, and $HXZ$ are indistinguishable even unambiguously since they are linearly dependent. Moreover, we have the following theorem.

   \begin{Theorem}
   	
   	Distinguishing the operators $X$, $Z$, $H$, and $HXZ$ (with errors) is as difficult as distinguishing the four BB84 states $|0\rangle$, $|1\rangle$, $|+\rangle$, $|-\rangle$.
   	
   \end{Theorem}

    The theorem could be verified as follow. Generally, to distinguish the four operators, Eve sends the B partita of a state $|s\rangle_{EB}\in C^{d}\otimes C^{2}$ to Bob and provides a bipartite measurement after Bob operated partita B. After Bob's operation, the problem for Eve is to distinguish $|s_{Z}\rangle=(I\otimes Z)|s\rangle$, $|s_{X}\rangle=(I\otimes X)|s\rangle$, $|s_{H}\rangle=(I\otimes H)|s\rangle$, $|s_{HXZ}\rangle=(I\otimes HXZ)|s\rangle$ in $C^{d}\otimes C^{2}$. On the other hand, to distinguish the four BB84 states, a protocol can be implemented as follow. An isometric transformation $U$ transforming $|0\rangle, |1\rangle\in C^{2}$ into $|s_{Z}\rangle, |s_{X}\rangle\in C^{d}\otimes C^{2}$, respectively, is implemented in the first step, which can be done since $|s_{Z}\rangle$ is orthogonal to $|s_{X}\rangle$, followed by a distinguishing protocol of the result states, namely $UT|0\rangle=|s_{T}\rangle$, where $T\in \{Z,X,H,HXZ\}$. Therefore, distinguishing the four BB84 states can be transformed into distinguishing the four operators while the error rate is not larger than the optimal error rate in distinguishing the operators. Similarly, by taking $|s\rangle=|0\rangle\in C^{2}$, distinguishing the four operators becomes the problem of distinguishing the four BB84 states. Therefore, distinguishing the four operators is as difficult as distinguishing the BB84 states.
    
    This result is not surprising since the dimension of the subspace of the four operators is 2 and thus no matter what state is employed to distinguish them, the four states would be in a subspace of dimension at most 2, in which at most two states can be distinguished. Note that increasing the dimension of the space would not give benefits for state distinguishing\cite{S2020The}.
    
    Hence, attacking Bob's bits would not provide benefits for Eve, compared with conventional attacks, namely eavesdropping on the channel, as in ordinary BB84 protocol.
    
    Therefore, the analysis of the above protocol can be justified as follow. By equivalently letting Bob provide measurements with untrusted measurement devices, the protocols are exactly the ordinary ones except for Bob's encoding operations. However, the encoding operation of Bob, namely operating a single qubit gate, could be nearly perfect. Note that the single qubit gates could be highly efficient in practice\cite{BD2008Efficient,MW2017Efficient}. Hence, the security and efficiency are nearly the same as the ordinary ones.
    
    \section{Employing more operators and the six-state type protocol}

   In the above protocol, Bob employs four operators. In fact, he can employ more, for example, eight operators which are $I$, $X$, $Z$, $XZ$, $H$, $HX$, $HZ$, and $HXZ$ for the BB84 type protocol. If the bases of Alice and Eve agree, Alice and Bob discard the states operated by the right four operators while if the bases of Alice and Eve do not agree, they discard the left four, see Table 1. The efficiency will not decrease but the operators of Bob could be more difficult to distinguish.

\begin{figure}
\includegraphics[width=1\textwidth]{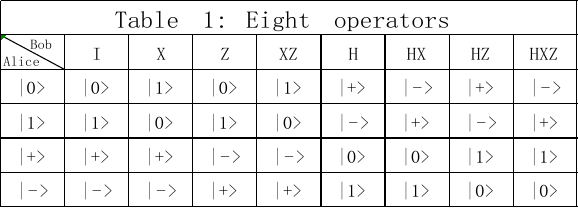}
\end{figure}
   	
    The same argument is suitable for employing three mutually unbiased bases as in the six-state protocol. In the measurement-device-independenization of the protocol, Alice randomly sends states in the three bases while Bob can employ twenty-four operators without decreasing efficiency, see Table 2, where
    $H_{1}=\frac{1}{\sqrt{2}}
    \begin{pmatrix}
    	1  &  -i\\
    	i  &  -1
    \end{pmatrix}$,
    $H_{2}=
    \begin{pmatrix}
    	1  &  0\\
    	0  &  i
    \end{pmatrix}$,
    $H_{3}=\frac{1}{\sqrt{2}}
    \begin{pmatrix}
    	1  &  1\\
    	i  &  -i
    \end{pmatrix}$,
    $H_{4}=\frac{1}{\sqrt{2}}
    \begin{pmatrix}
    	1  &  -i\\
    	1  &  i
    \end{pmatrix}$,
    $|a\rangle=\frac{1}{\sqrt{2}}(|0\rangle+i|1\rangle)$, $|b\rangle=\frac{1}{\sqrt{2}}(|0\rangle-i|1\rangle)$, and all states are up to global phases which affect nothing.

    Hence, only $\frac{1}{3}$ of the states are left as in the six-state protocol. For example, if Alice employs basis $\{|0\rangle, |1\rangle\}$ while Eve measures via $\{|0\rangle, |1\rangle\}$, then states on which Bob operates $I$, $X$, $Z$, $XZ$, $H_{2}$, $H_{2}X$, $H_{2}Z$, $H_{2}XZ$ are left with others discarded. Here, Alice and Bob can decide on coding methods after Eve's turn and error estimating, namely, for example, they could agree to encode 0 by $I, Z, H_{2}, H_{2}Z$ and encode 1 by $X, XZ, H_{2}X, H_{2}XZ$ after estimating error rate by states in randomly chosen operators. See Table 2.

\begin{figure}
		\includegraphics[width=1.0\textwidth]{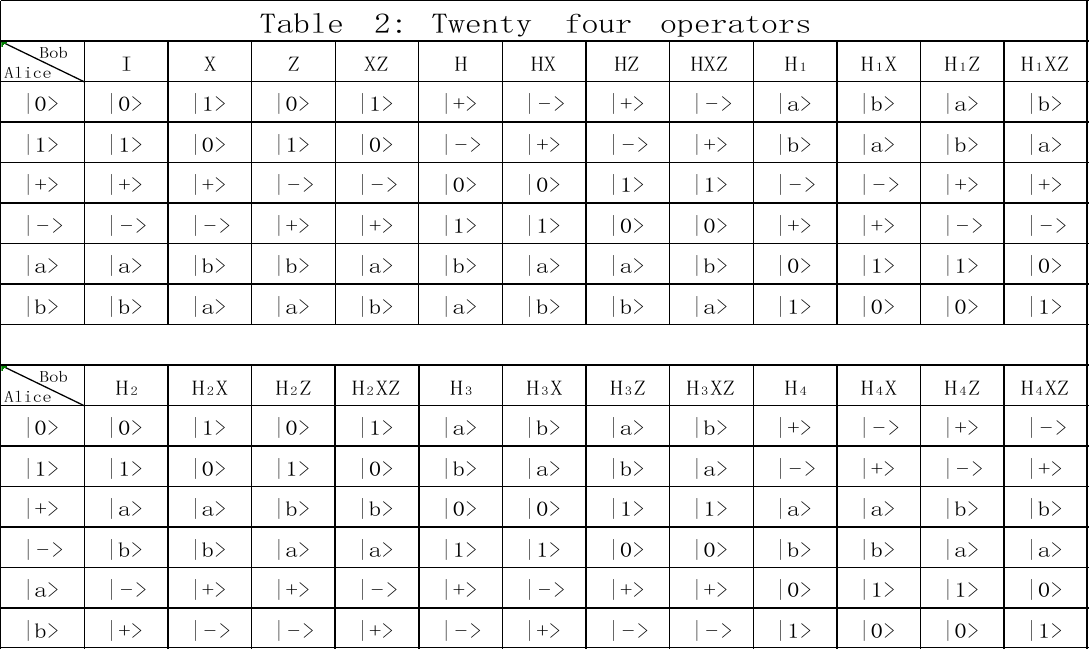}
\end{figure}
    
	\section{Measurement-device-independenization of prepare-measure QKD protocols}
	
	Let us investigate the general scheme of measurement-device-independenization for a prepare-measure QKD protocol. Assume that the ordinary protocol encodes 0 by state $|0\rangle$ and 1 by state $|y\rangle=x_{1}|0\rangle-x_{0}|1\rangle\neq |0\rangle$, where without loss generality $x_{0}, x_{1}$ are real and hence $x_{0}^{2}+x_{1}^{2}=1$ since we can always view the two vectors in the same plane. By making the encoding states symmetric, namely, the bit 0 is encoded by states $|0\rangle$ and $|x\rangle=x_{0}|0\rangle+x_{1}|1\rangle$ while the bit 1 is encoded by states $|1\rangle$ and $|y\rangle=x_{1}|0\rangle-x_{0}|1\rangle$, respectively, the protocol is described as follow.
	
    For each bit, Alice randomly sends one of the four states to Bob. After Bob receives the state, he implements the PNP scheme and randomly provides an operator in one of $I$, $U$, $XZ$, $UXZ$, $X$, $Z$, $UX$, $UZ$, $XU$, $ZU$, $UXU$ and $UZU$ to the state employed for coding, where
    $U=\begin{pmatrix}
    	x_{0}  &  x_{1}\\
    	x_{1}  &  -x_{0}
    \end{pmatrix}$
    is the (real) unitary operator transforming basis $\{|0\rangle, |1\rangle\}$ to basis $\{|x\rangle, |y\rangle\}$. Then Bob sends the state together with a chosen basis to Eve for measuring. In detail, Bob chooses basis $\{|0\rangle, |1\rangle\}$ if he operates $X$, $Z$, $XU$ or $ZU$, chooses basis $\{|x\rangle, |y\rangle\}$ if operates $UX$, $UZ$, $UXU$, or $UZU$, while randomly chooses one of the two bases if he operates others. For Eve, she is required to measure the state via the chosen basis and announce the outcome. Finally, Alice announces her basis choices for Bob to discard those bits with incompatible bases while others are employed for estimating the bit error rate and generating a raw key as in the BB84 type protocol.

    In such a protocol, $\frac{1}{3}\times\frac{1}{2}+\frac{1}{3}\times 1+ \frac{1}{3}\times 0=\frac{1}{2}$ states would be left after the sifting procedure, see Table 3, where N represents that the pair of Alice's state and Bob's operator is invalid no matter what bases does Bob choose. In fact, the measurement bases can be chosen by Eve but the efficiency would be halved since the bits with non-matched bases should be discarded.

\begin{figure}
	\includegraphics[width=1\textwidth]{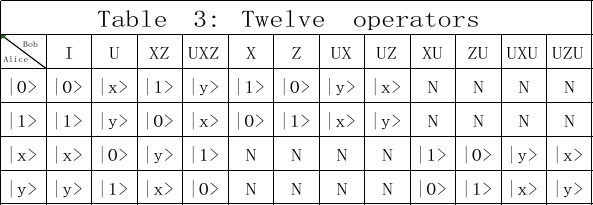}
\end{figure}

    The discussions of protocols employing more than two bases are similar. It is worth noting that the efficiency of a general protocol might not be higher than the BB84 type protocol, which employs mutually unbiased bases, while the optimal strategy for the eavesdroppers and thus the secure bounds for the legitimate partner might need to recalculate. Despiteness, the theoretical feasibility of the scheme for coding by general states is verified. Most of the prepare-measure QKD protocols can be measurement-device-independenized by the scheme. It might not be surprising that the scheme can be modified for entanglement-based protocols.

    \section{Photon-number-adding attack, Photon number purification scheme, and practical implement}

    In this section, we will investigate the photon-number-adding(PNA) attack, which is exactly the Trojan horse attack when applied in DDI protocols, and provide a solution called the photon-number-purification(PNP) scheme. Then we will modify the BB84-type protocol for practical settings.

    \subsection{Photon-number-adding attack: A problem}

    The PNA attack is described as follows. Eve adds several states in the signal (state) Alice sends to Bob (or even substitutes the signal by hers). These states are employed for distinguishing Bob's operators and will be removed before measuring. Note that despite the indistinguishability of Bob's operators via a single state, they can be unambiguously distinguished via multiplied copies. If Bob can not count the photon numbers in a signal, Eve can implement such an attack without being noticed. Essentially, it is such an attack that breaks the security of certain protocols such as the detector-devices-independent(DDI) QKD protocol\cite{LK2014Detector,Q2015Trustworthiness,GR2015Quantum,LL2015Simple,SH2016Insecurity}, the protocol in \cite{H2021Measurement}, and the plug-and-play protocol\cite{MH1997Plug,GF2006Trojan}.
    
    \subsection{Photon number purification: A solution}
    
    The PNA attack can be handled by applying C-NOT gates which can be highly efficient, namely with more than $99\%$ fidelity\cite{KW2021Demonstration,NT2022Fast}. We provide a solution for BB84-type protocol while it would be easily generalized to others. Denote the two C-NOT gates as follow, where $x,y$ denote the partite.
    
   \begin{small}
    \noindent$\begin{matrix}
    	C_{0_{xy}}: &C^{2}\otimes C^{2} \rightarrow  C^{2}\otimes C^{2}\\
    	&|0\rangle_{x}|0\rangle_{y} \rightarrow|0\rangle_{x}|0\rangle_{y}\\ 	
    	&|1\rangle_{x}|0\rangle_{y} \rightarrow|1\rangle_{x}|1\rangle_{y}
    \end{matrix}$\quad
$\begin{matrix}
	C_{+_{xy}}:&C^{2}\otimes C^{2} \rightarrow  C^{2}\otimes C^{2}\\
	&|+\rangle_{x}|+\rangle_{y} \rightarrow  |+\rangle_{x}|+\rangle_{y}\\ 	
	&|-\rangle_{x}|+\rangle_{y} \rightarrow  |-\rangle_{x}|-\rangle_{y}
\end{matrix}$
\end{small}

    The scheme is described as follows. Before Bob encodes the state sent by Alice, he randomly copies the state via basis $\{|0\rangle,|1\rangle\}$ or $\{|+\rangle,|-\rangle\}$, namely, employs source $|0\rangle_{B}$ and operates $C_{0_{AB}}$ or employs source $|+\rangle_{B}$ and operates $C_{+_{AB}}$, where $B$ denotes the auxiliary partita and $A$ denotes the ordinary partita. Then he aborts the ordinary state which might still be employed for estimating, and employs partita $B$. Hence, extra states in the signal are removed and the PNA attack becomes a substituted attack (which might even not be successful), for which the security is guaranteed. We will call the procedure $'photon\ number\ purification'$(PNP). After it, Bob encodes partita $B$ as usual and sends it to measure.
    
    In such a procedure, a state is discarded if it is wrongly purified, namely, if the state sent by Alice is not on the same basis chosen by Bob for copying. The same investigations are suitable for the six-state type protocol, in which Bob purifies via randomly chosen three bases.

    Note that the PNP scheme could also be applied to DDI-QKD protocol, in which the PNA attack is exactly the combination of an intercept-resend attack and a Trojan horse attack\cite{LK2014Detector,Q2015Trustworthiness,GR2015Quantum,LL2015Simple,SH2016Insecurity}, as well as the protocol in \cite{H2021Measurement} or plug-and-play protocol, to close the loopholes\cite{MH1997Plug,GF2006Trojan}. To verify this, just let the legitimate partner implement the PNP scheme after receiving signals and before further operations.
    
    In fact, the PNP scheme might be viewed as a kind of device isolation and thus might be applied to other tasks too, besides QKD. The input signal will not be output while the output signal does not contain the input photons. The same work can be done by teleportation\cite{BB1993Teleporting}, but the PNP scheme could be more efficient since it does not need to employ entanglement and Bell measurement.

    \subsection{The practical BB84 type protocol}

    Inspirited by the above investigations, the final BB84 type protocol employing weak coherence sources and without photon number counting technologies is described as follows, see Figure 2.

\begin{figure}
	\includegraphics[width=1.0\textwidth]{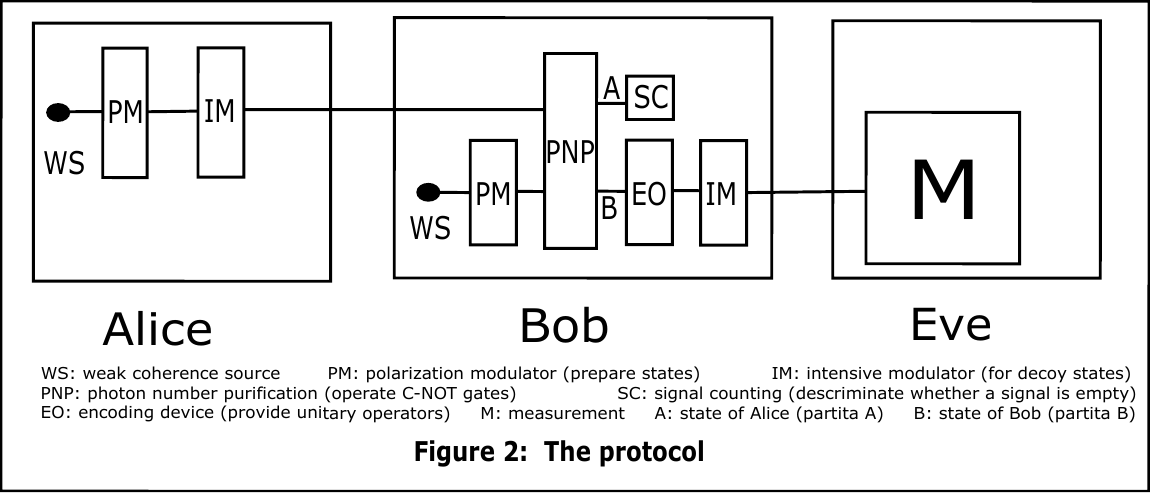}
\end{figure}

    In detail, Alice prepares and sends her qubit to Bob with the decoy-state method\cite{H2003Quantum,LM2005Decoy,MQ2005Practical} against photon-number-splitting(PNS) attack\cite{HI1995Quantum,LJ2002Quantum} (step 1) while Bob employs the PNP procedure above. The ordinary signal (partita A) is employed for estimating the channel between Alice and Bob, for which Bob only needs to determine whether it is empty with insecure devices, making the decoy-state methods of Alice work. Then Bob encodes the signal of partita B and sends it (also by the decoy-state method) to untrusted Eve for measuring (steps 2, 3). Steps 4 and 5 are similar to the protocol with ideal assumptions.
    
    Note that in the protocol, all non-decoy states sent by Alice are employed by Bob. Therefore, to generate a raw secret bit, 4 qubits needed to be sent by Alice, not including the decoy ones, of which half are employed as decoy states by Bob, while in the scenario that Bob assumes the measurements instead of Eve, only 2 qubits needed to be sent. Note that in the standard MDI-QKD protocol\cite{LC2012Measurement} (employing Bell measurements with two inclusive results), 8 qubits, not including the decoy ones, are required for obtaining a secret bit. Therefore, our protocol could even be more efficient.
    
	\section{Conclusion}

	In conclusion, we presented MDI-QKD protocols without employing joint measurements. Their performances can nearly be the same as the ordinary ones (for example, BB84 protocol) since they can be analyzed in exactly the same way, except ours are measurement-device-independent, assisted by the PNP scheme which could be nearly perfect. Employing more bases or operators could improve their security, as we have discussed in the situation of the six-state protocol, while the legitimate partner can decide the coding methods even after the turns of Eve and error estimating. The scheme is suitable for general prepare-measure QKD protocols, making them MDI-QKD protocols, which is called $'measurement-device-indepenization'$ of QKD protocols. It will not be surprising that it can be modified for even entanglement-based protocols.
	
	We also investigated the PNA attack and presented the PNP scheme as a solution, which might also be employed in other tasks. Finally, we provided the BB84-type protocol in practical settings. The protocol could be more efficient than the standard MDI-QKD protocol while it does not need to employ joint measurements.
	
	Although our protocol seems similar to the DDI-QKD protocol which has been proven to be insecure under certain attacks, they are different both in the need for joint measurements and the need for security in measurement devices. The loopholes in the DDI-QKD protocol are not suitable for ours since the PNP scheme. In fact, the PNP scheme can also be employed for closing the loopholes in DDI protocol as well as two-stage protocols such as plug-and-play protocol.
	
	Finally, there might be a corresponding version for the asymptotically optimal protocols\cite{S2021Asymptotically} which might not be efficient in practice, however, might provide an asymptotic secure bound of the bit error rate.

    \bibliographystyle{unsrt}
    \bibliography{Bibliog}

	\end{document}